\documentclass[11pt]{article}

\usepackage{amsmath}
\usepackage{amssymb}
\usepackage{fullpage}
\usepackage{graphicx}
\usepackage{bm}% bold math
\usepackage{hyperref}% add hypertext capabilities
\usepackage{cite}
\usepackage{authblk}
\usepackage{lineno}
\usepackage{xcolor}
\PassOptionsToPackage{hyphens}{url}\usepackage{hyperref}

\title{\textbf{Collective Movement with Signaling}}
\date{}

\author{\textbf{Mohammad Salahshour}\thanks{\texttt{salahshour.mohammad@gmail.com}.}}
\affil{Max Planck Institute for Mathematics in the Sciences, Inselstrasse 22, D-04103, Leipzig, Germany}
\author{\textbf{Shahin Rouhani}\thanks{\texttt{srouhani@sharif.ir}.}}
\affil[2]{Department of Physics, Sharif University of Technology, P.O. Box 11165-9161, Tehran, Iran}

\begin{document}
\maketitle

\begin{abstract}
	We consider a population of mobile agents able to make noisy observation of the environment and communicate their observation by production and comprehension of signals. Individuals try to align their movement direction with their neighbors. Besides, they try to collectively find and travel towards an environmental direction. We show that, when the fraction of informed individuals is small, by increasing the noise in communication, similarly to the Viscek model, the model shows a discontinuous order-disorder transition with strong finite size effects. In contrast, for large fraction of informed individuals, it is possible to go from the ordered phase to the disordered phase without passing any phase transition. The ordered phase is composed of two phases separated by a discontinuous transition. Informed collective motion, in which the population collectively infers the correct environmental direction, occurs for high fraction of informed individuals. When the fraction of informed individuals is low, misinformed collective motion, where the population fails to find the environmental direction becomes stable as well. Besides, we show that an amount of noise in the production of signals is more detrimental for the inference capability of the population, and increases the density fluctuations and the probability of group fragmentation, compared to the same amount of noise in the comprehension.
\end{abstract}

	\noindent
	
	\section*{Introduction}
	
	Many species, from bacteria \cite{Sokolov,Czirok} and cells \cite{Szabo,Friedl} to insects \cite{Buhl,Couzin1}, large animals \cite{Ward,Bajec,Fischhoff,Sueur} and humans \cite{Faria} show collective motion, an intriguing phenomena in which individuals in a population move in ordered groups, presumably due to local interactions \cite{Vicsek}. Such a collective motion is suggested to endow many advantages, such as avoiding predators \cite{Ward2}, or enhancing the information acquisition capability of the population in noisy environments \cite{Grunbaum,Berdahl}. Although collective motion has been subject to intense study \cite{Vicsek,Vicsek2,Couzin,Chate,Gregoire,Cucker,Bricard,Nagai,Luca}, an important point not considered in the relevant literature, is that in many cases, the information individuals reach from others in the population is provided through communication by exchanging signals \cite{Friedl,McCann,Haas,Rappel,Leonhardt,Sumpter}. This raises the important question that how collective motion in a population of individuals who exchange their social information by production and comprehension of signals is formed, and how the noise inherent in the communication system affects the collective information acquisition capability of the moving population? In addition to the biological examples, questions can arise in a community of artificial agents who are capable of making noisy communication \cite{Monaco1,Monaco2}. To answer these questions, we consider a population of mobile agents trying to collectively find and travel towards an environmental direction. Individuals communicate information about their movement by production and comprehension of signals and possibly make noisy observations about the environmental direction. They make decisions about their direction of motion based on this information. We show that \textit{informed collective motion}, in which the population collectively moves towards the environmental direction, emerges if the noise in communication is low enough and the fraction of informed individuals, who are able to make noisy observation of the environment, is high enough. As the fraction of informed individuals decreases (for small communication noise), the model shows a discontinuous phase transition to a \textit{misinformed collective motion} phase in which the population moves collectively, but towards a direction other than the environmental direction. By increasing the noise in communication, for a low fraction of informed individuals, the model shows a discontinuous phase transition to a disordered phase in which individuals move towards random directions. As is the case in Vicsek-like models \cite{Vicsek2,Chate,Gregoire}, the order-disorder transition suffers from strong finite-size effects that make its discontinuous nature apparent only in large system sizes. On the other hand, when the fraction of informed individuals is high, such that the system is well into the informed collective motion phase, by increasing the communication noise, contrary to the Viscek-like models, the population moves gradually from the ordered phase to the disordered phase without any phase transition. This shows how the amount of information about the environment contained in the population can change the nature of the order-disorder transition observed in Vicsek-like models.
	
	Finally, we show an amount of noise in the production of signals is more detrimental for the collective information acquisition capability of the population compared to the same amount of noise in the comprehension of signals. This result is in keeping with a recently found comprehension-production asymmetry in a model of collective decision making, where individuals residing on a network, try to form a consensus about an environment which can be found in a finite number of possible states \cite{Salahshour2}. Besides, we show that the production noise increases the density fluctuations and the probability of group fragmentation, compared to comprehension noise. Thus, in this regard, by extending results found in ref. \cite{Salahshour2}, our findings suggest asymmetry between signal comprehension and production is a fundamental characteristic of biological communication systems.
	
	\begin{figure}[!ht]
		\centering
		\includegraphics[width=1\linewidth, trim = 15 24 15 12, clip,]{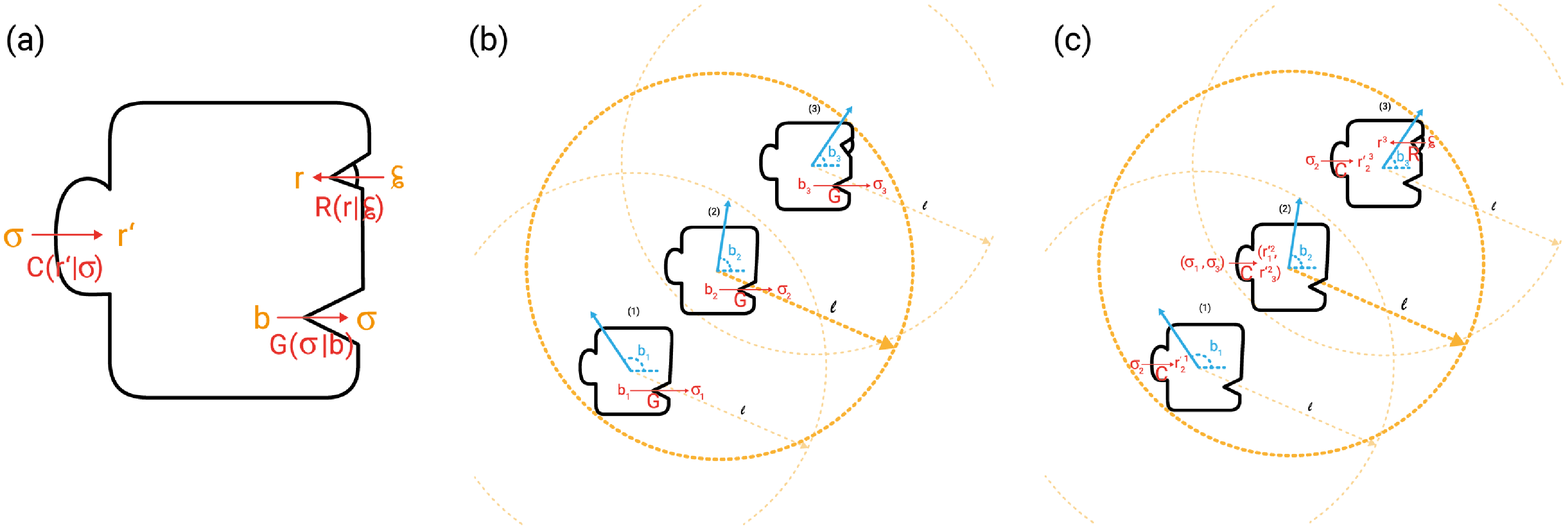}
		\caption{Schematic representation of the model. (a): Each agent is equipped with two information channels to communicate. Production channel $G(\sigma|b)$ is used for signal production, such that signal $\sigma$ is produced for a direction $b$ ‌with probability$G(\sigma|b)$. In the same way, Comprehension channel $C(r'|\sigma)$, is used for comprehension of the signals. In addition to these, a fraction $q$ of the individuals are informed individuals, who are equipped with an observation channel $R(r|\epsilon)$. (b) and (c): The dynamics of the model. Agents move on a two-dimensional space with constant speed $v_0$ and towards different directions $b_i$. In each time step, each agent $i$ produces a signal $\sigma_i$ based on its direction $b_i$ to communicate its direction of motion (b). The signal travels up to a distance $l$, such that all the agents up to a distance $l$ from a transmitter receive its signal. Receivers comprehend signals using their comprehension channel (c). Here, agents 1 and 3 receive a signal only from agent 2, but agent 2, having two individuals in its $l$-neighborhood, receive signals from both $1$ ‌and$3$. Besides, agent 3, who is an informed individual makes an observation using its observation channel. Each agent $i$, makes a decision based on the information they receive from communication $\pmb{r'}^i$, and the result of observation $r^i$ (in the case of informed individuals), using the weighted averaging rule eq. \eqref{eq1} . Here, $\pmb{r'}^1=(r'^1_2)$, $\pmb{r'}^2=(r'^2_1,r'^2_3)$, $\pmb{r'}^3=(r'^3_2)$.
		}
		\label{fig0}
	\end{figure}
	\section*{The Model}
	
	A schematic representation of the model is provided in FIg. (\ref{fig0}). We consider a population of $N$ mobile agents, moving with constant speed $v_0$ on a $L\times L$ two dimensional surface with periodic boundaries. There is a favorable direction of motion $\epsilon$, called the environmental direction. Individuals try to find and travel towards $\epsilon$. Individuals decide about the direction of their motion based on their personal observation (if available) and social information. Social information is acquired by production and comprehension of signals. To communicate its direction of motion $b$, each individual produces a signal $\sigma$, using its signal production channel $G(\sigma|r=b)$. That is a signal in the interval $[\sigma-d\sigma/2,\sigma+d\sigma/2]$ is produced when an individual intends to signal direction $b$ with probability $G(\sigma|r=b)d\sigma$. Here, $d\sigma$ is a differential element, and $b$ and $\sigma$, referring to a direction in $2$ dimensions, satisfy $b,\sigma\in[0,2\pi)$. Signals are transmitted up to a distance $l$. That is, all the individuals in a circle of radius $l$ centered around the transmitter receive the signal. The receivers, comprehend a signal to refer to a direction of travel $r'\in[0,2\pi)$, according to their comprehension channel $C(r'|\sigma)$. That is, signal $\sigma$ is comprehended as referring to a direction in the interval $[r'-dr'/2,r'+dr'/2]$ with probability $C(r'|\sigma)dr'$. In general, signal production and comprehension are subject to noise \cite{Brumm,Hotchkin,Wiley,Schuster,Salahshour2,Salahshour}. To implement this fact, we take the production $G(\sigma|b)$ and comprehension $C(r'|\sigma)$ channels to be uniform distributions in the intervals, respectively, $[b-\eta_G,b+\eta_G]$ and $[\sigma-\eta_C,\sigma+\eta_C]$. This is done by taking $\sigma=b+\xi_G$ and $r'=\sigma+\xi_C$, where $\xi_G$ and $\xi_C$ are uniformly distributed random numbers in the intervals, respectively, $[-\eta_G,\eta_G]$ and $[-\eta_C,\eta_C]$. $\eta_G$ and $\eta_C$ can be thought of as the noise level, in respectively, production and comprehension of signals.
	
	Personal information is acquired by noisy observation of the environment. We assume observation is made through a noisy channel $R(r|\epsilon)$, such that the result of an observation in environment $\epsilon$ is in the interval $[r-dr/2,r+dr/2]$ with probability $R(r|\epsilon)dr$. We take $R(r|\epsilon)$ to be uniformly distributed in the interval $[\epsilon-\eta_R,\epsilon+\eta_R]$. $\eta_R$ can be thought of as the noise level in observation. We assume only a fraction $h$ of the individuals, called informed individuals, are able to observe the environment.
	
	As a result of signals an individual receives, it reaches a set of representations $\pmb{r'}$. This set is composed of all the directions an individual receives by comprehending signals in its $l$ neighborhood. Besides, informed individuals make a personal observation, $r$. Each individual, $\alpha$, makes a decision about its direction of motion $b_\alpha$, based on its observation $r_\alpha$ (in the case of informed individuals) and social information $\pmb{r'}_\alpha$, by a weighted averaging rule. That is:
	\begin{align}
	b_{\alpha}=\omega r_{\alpha}+(1-\omega)\frac{\sum_{r'\in \pmb{r'}_\alpha}r'}{|\pmb{r'}_\alpha|}.
	\label{eq1}
	\end{align}
	Here, $|\pmb{r'}_\alpha|$ is the number of representations (directions) individual $\alpha$ has received, and $\omega$ is the self-confidence of individuals. $\omega$ lies between $0$ and $1$ and determines how individuals weigh their personal observation compared to their social information in decision making.
	
	As in the Vicsek model \cite{Vicsek2}, the dynamic is synchronous. That is at each time step, informed individuals make an observation using $R(r|\epsilon)$, and all the individuals transmit a signal $\sigma$ based on their direction of motion $b$ using $G(\sigma|b)$. The signals are received by all the individuals up to a distance $l$ of the transmitter. Receivers, comprehend the signals as referring to direction $r'$ using $C(r'|\sigma)$. Finally, individuals make a decision about their direction of travel using their decision-making rule and update their direction accordingly. The simulations start with a random distribution of the position and direction of individuals. The base parameter values used in the simulations (unless otherwise stated) are $l=1$, $L=10$, $N=100$, $v_0=0.1$, $\eta_R=0.75$, $\omega=0.25$.
	
	In the following we use two variables to distinguish different phases of the system. These are the absolute value of the normalized average velocity of the population $m=|\sum_{\alpha=1}^{N}\frac{\vec{v}_\alpha}{Nv_0}|$ and the angular deviation of the average population direction from the environmental direction $\delta\theta=\sum_{\alpha=1}^{N}\frac{\theta_\alpha-\epsilon}{N}$. For a population with observation noise, production noise, and comprehension noise, equal respectively to, $\eta_R$, $\eta_G$ and $\eta_C$, and with a fraction $h$ of informed individuals, we show these by, respectively, $m(\eta_R,\eta_G,\eta_C,h)$ and $\delta\theta(\eta_R,\eta_G,\eta_C,h)$.
	
	\begin{figure}%[ht]
		\centering
		\includegraphics[width=1\linewidth, trim = 105 4 95 7, clip,]{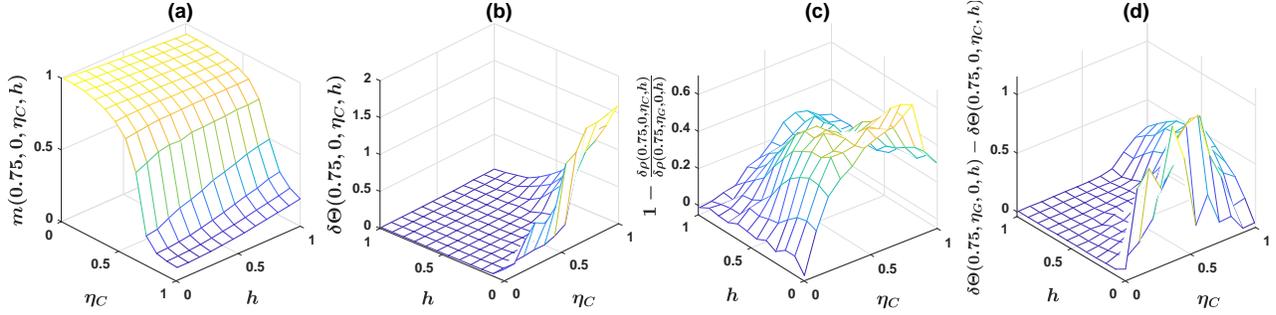}
		\caption{(a) and (b): $m(0.75,0,\eta,h)$ and $\delta\Theta(0.75,0,\eta,h)$ in the $\eta-h$ plane. For low communication noise $\eta$, collective motion emerges. As noise increases, the system goes to the disordered phase. The collective motion phase is composed of an \textit{informed collective motion} phase, when the fraction of informed individuals $h$ is large, and a \textit{misinformed collective motion} phase, when $h$ is small. (c) and (d): Asymmetry in density fluctuations $1-\frac{\delta\rho(0.75,0,\eta,h)}{0.75,\eta,0,h}$ (c), and the difference in angular deviation from the environmental direction when an amount of noise $\eta$ is in production with that when the noise is in comprehension  (d) in the $\eta-h$ plane. Production noise increases density fluctuations and deters the inference capability of the population. Here, $L=10$, $v_0=0.1$, $\omega=0.25$, $\eta_R=0.75$ and $\rho=1$. The simulation is run for $T=4000$ time steps, and an average over the last $500$ time steps and $R=24$ runs is taken.}
		\label{fig1}
	\end{figure}
	\begin{figure}[ht]
		\centering
		\includegraphics[width=1\linewidth, trim = 55 82 170 50, clip,]{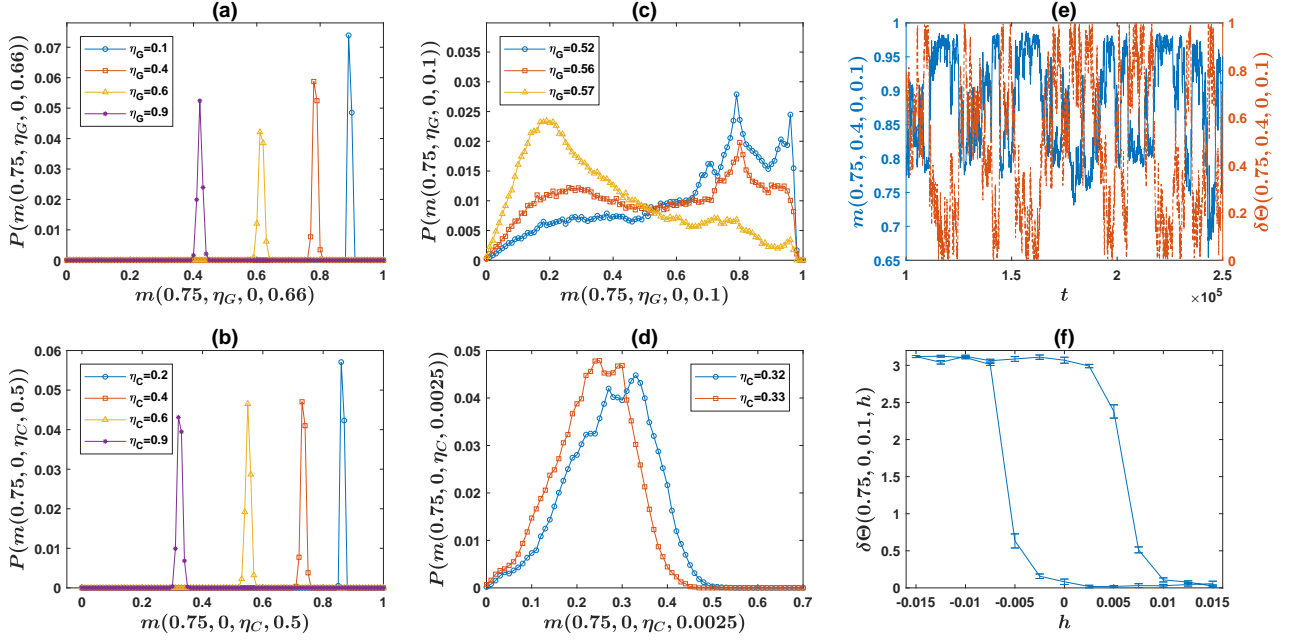}
		\caption{(a) and (b): For large $h$, the order parameter decrease gradually as noise in communication increases, for both production (a) and comprehension (b) noise. Here $L=300$, $v_0=0.6$, and $\rho=0.1$. (c) and (d): For small $h$, where misinformed collective motion is possible, the order-disorder transition is discontinuous for both production (c) and comprehension (d) noise. In (c) $L=100$, $v_0=0.6$, and $\rho=0.1$ and in (d) $L=200$, $v_0=0.1$, and $\rho=0.1$. (e): $m(0.75,0.4,0,0.1)$ (blue solid lines) and $\delta\Theta(0.75,0.4,0,0.1)$ (red dashed line) as a function of time. With production noise the population is decomposed into several dense independently moving groups. The evolution shows intermittency between the informed collective motion, where different groups move towards the environmental direction (high $m$ low $\delta\Theta$) and misinformed collective motion where different groups head towards different directions (low $m$ high $\delta\Theta$). Here, $L=100$, $v_0=0.6$, and $\rho=0.1$. (f): Hysteresis loop shows the informed-misinformed collective motion transition is of first order. Here, $L=20$, $v_0=0.1$, and $\rho=1$. In all the simulations $\omega=0.25$ and $\eta_R=0.75$.}
		\label{fig2}
	\end{figure}
	\section*{Results}
	\subsection*{phase transitions}
	In Fig. (\ref{fig1}.a), the absolute value of the normalized average velocity of the population in the case that an amount of noise $\eta$ is in the comprehension and production is noiseless, $m(0.75,0,\eta,0.1)$ is plotted. The case that the noise is in production is qualitatively similar. For low noise level, $m$ is close to $1$. This shows that collective motion emerges, and individuals travel in the same direction. By increasing the noise level, the model shows a transition to a disordered phase in which individuals fail to align their motion and travel to random directions. Thus, $m$ takes a small value. Interestingly, the ordered phase is composed of two distinct phases. To see this, in Fig. (\ref{fig1}.b), we plot the angular deviation of the average direction of motion from the environmental direction, $\delta\theta(0.75,0,\eta,0.1)$. In the ordered phase, for large $h$, the population possesses enough informed individuals to be able to collectively find the environmental direction. Consequently, $\delta\theta(0.75,0,\eta,0.1)$ takes a value close to zero. We call this phase \textit{informed collective motion} phase. On the other hand, for small $h$, misinformed collective motion, in which the population moves collectively in a wrong direction, becomes possible as well, and the system becomes bistable (see below). Consequently, $\delta\theta(0.75,0,\eta_C,0.1)$ takes a large value. We call this phase \textit{misinformed collective motion} phase.
	
	The nature of the order-disorder transition in the Vicsek and related models has been the subject of intense research \cite{Vicsek,Gregoire,Vicsek2,Chate}. It is well known that the order-disorder transition is discontinuous in many cases (depending on the parameter values and the way noise is implemented in the model) \cite{Vicsek}. However, due to strong finite-size effects, the discontinuous nature of the transition shows up only in very large sizes \cite{Chate}. Our model shows a similar phenomenology only for small $h$. However, the situation is different for large $h$. This can be seen in Fig. (\ref{fig2}.a) and Fig. (\ref{fig2}.b) for respectively, the production and comprehension noise, where the probability distribution of the order parameter for different noise levels and constant (large) $h$ is plotted. Here, the distribution is derived from a single time series of the system of length $T=50000$, after discarding the first $1000$ time steps. As can be seen, by increasing the noise level, the order parameter gradually decreases. This excludes a discontinuous transition. Besides, the distribution remains peaked at a single value and no broadening of the distribution resulting from large fluctuations characteristic of a continuous transition occurs \cite{Goldenfeld}. This suggests it is possible to go from the informed collective motion phase to the disordered phase without passing any phase transition. 
	
	In Fig. (\ref{fig2}.c) and Fig. (\ref{fig2}.d), by plotting the distribution of the order parameter for fixed $h$ and different noise levels, we confirm the discontinuous nature of the order-disorder transition for small $h$. In Fig. (\ref{fig2}.c) the case of production noise is considered. Here, $L=100$, $v_0=0.6$, $\omega=0.25$, $\eta_R=0.75$ and the density $\rho=0.1$. The distributions are derived from a single time series of the system of size $T=1.2\times 10^6$, after discarding the first $10^5$ time steps. As can be seen, the distribution shows distinct peaks corresponding to the ordered and disordered phases. As the noise level increases, the peaks corresponding to the ordered phase decrease, while that corresponding to the disordered phase increases. This phenomenology is characteristic of a discontinuous transition \cite{Binder,Chate}. In Fig. (\ref{fig2}.d), the case of comprehension noise is considered, where the same bi-modality which suggests a discontinuous transition is observed. Here, $L=200$, $v_0=0.1$, $\omega=0.25$, $\eta_R=0.75$ and $\rho=0.1$. We note that the finite-size effects are much stronger for comprehension noise compared to production noise, such that the discontinuous nature of the transition becomes apparent for larger population size in the former. Besides, for production noise, the finite-size effects are stronger for smaller velocities, while for comprehension noise, they are stronger for larger velocities. We have not been able to conclusively infer the discontinuous nature of the transition for large velocities when noise is in the comprehension.
	
	Returning to Fig. (\ref{fig2}.a), we see that for the case of production noise the distribution of the order parameter has two major peaks in the ordered phase. This happens only for small enough $h$, i.e. when the ordered phase is bistable and the system can be found in both informed and misinformed collective motion phases. The two major peaks of $m$ correspond to these two phases. This can be seen in Fig. (\ref{fig2}.e), where $m$ together with $\delta\Theta$, as a function of time, are plotted. As can be seen, $m$ shows intermittency between two values corresponding to the two peaks of the ordered phase in Fig. (\ref{fig2}.c). When $m$ is very large, corresponding to the rightmost peak in Fig. (\ref{fig2}.c), $\delta\Theta$ is very small, indicating that the average direction of motion coincides with the environmental direction. On the other hand, when $m$ takes the smaller value, $\delta\Theta$ becomes large, indicating the average population direction differs from the environmental direction. The reason why the value of $m$ in the misinformed collective motion phase is smaller than that in the informed collective motion phase is that when noise is in production (as is the case here) the probability of group fragmentation is high, such that in many times, the population is composed of different groups each collectively moving towards a different direction independently of others. Occasionally a group is decomposed into smaller groups, or different groups can merge to form a larger group [see the Supplementary Video for a visual manifestation]. In the informed collective motion phase, all the groups head towards the environmental direction. Thus $m$ takes the largest value. While in the misinformed phase, different groups can head in different directions. This decreases $m$ as Fig. (\ref{fig2}.c) suggests.
	
	For comprehension noise, $m$ does not show a similar bi-modality that indicates intermittency between informed and misinformed collective motion. The reason is that, contrary to the production noise, with comprehension noise, the population rarely is decomposed into different dense groups with different direction of travel, and the strong fission-fusion dynamics observed for production noise is absent in the case of comprehension noise [see the Supplementary Video]. We will shortly return to this difference between comprehension and production noise.
	
	The bi-stability associated with a discontinuous transition results in hysteresis, which provides an alternative way to test the nature of the informed-misinformed phase transition \cite{Binder}. This is shown in Fig. (\ref{fig2}.f), where the hysteresis loop for the case of comprehension noise is shown. Production noise shows similar hysteresis effects. Here, we run a simulation beginning with $h=\frac{6}{400}$, which lies in the informed consensus phase. We gradually decrease $h$ down to $h=-\frac{6}{400}$ (a negative $h$ results from reversing the environmental direction) and then increase it back to the initial value. The resulting hysteresis loop results from the memory effects and indicates a discontinuous transition.
	
	\subsection*{Noise in signal production increases the density fluctuations and degrades collective information acquisition capability} We have already seen that production noise increases the probability of group fragmentation and leads to a strong fission-fusion dynamic absent for the comprehension noise. This can be shown more quantitatively. For this purpose, we define the (relative) asymmetry in density fluctuations as $\frac{\delta\rho(\eta_R,\eta,0,h)-\delta\rho(\eta_R,0,\eta,h)}{\delta\rho(\eta_R,\eta,0,h)}$, where, the density fluctuation $\delta\rho$, is defined as the standard deviation of spatial density of the population. In a situation where the individuals are distributed uniformly in space, which is often the case for comprehension noise, this takes a small value. On the other hand, when the group is decomposed into independently traveling dense groups, as it is often the case for the production noise, this takes a large value. Consequently, the asymmetry in density fluctuations is always positive, as can be seen in Fig. (\ref{fig1}.c). This shows, compared to comprehension noise, the production noise increases the density fluctuations and the probability of group fragmentation.
	
	There is another asymmetry between comprehension and production of signals. This asymmetry arises in the collective information acquisition capability of the population. To see this, we define the asymmetry in the inference capability of the population as the difference between the angular deviation of the average population direction from the environmental direction when an amount of noise is in signal production compared to the case when the noise is in signal comprehension, $\delta\Theta(\eta_R,\eta,0,h)-\delta\Theta(\eta_R,0,\eta,h)$. To justify this choice, we note that in the case that production noise is more detrimental for the collective information acquisition of the population, it leads to a higher angular deviation compared to the case that the same amount of noise is in the comprehension. Consequently, this quantity becomes positive. The asymmetry in the inference capability of the population, $\delta\Theta(\eta_R,\eta,0,h)-\delta\Theta(\eta_R,0,\eta,h)$, for $\eta_R=0.75$, in the $\eta-h$ plane is plotted in Fig. (\ref{fig1}.d). As can be seen, this quantity is always positive. This indicates that, compared to the same amount of error in the comprehension, an amount of error in production leads to a poor collective inference, and thus, a larger angular deviation from the environmental direction. We note that the positivity of this quantity results from two shifts in the phase transitions. First production noise shifts the order-disorder transition to smaller noise levels. This shows production noise is more detrimental to the ordering of the population. Second, production noise shifts the monostable informed collective motion phase to larger values of $h$. This means with production noise, a larger fraction of informed individuals is necessary for the population to successfully infer the correct environmental direction. As shown in the Supplementary Material, both asymmetries are robust features of the model, valid for all the parameter values.
	
	\section*{Discussion}
	We have introduced a model of collective movement in which individuals in a population try to collectively find and travel to a preferred direction. To do so, individuals can use social information provided by the noisy observation of the environment, and social information, provided by communication between individuals by exchanging signals. By analysis of the model, we showed that production noise not only decreases the inference capability of the population but also increases density fluctuations and the probability of group fragmentation. This suggests devoting resources to noise reduction in signal production is more beneficial than noise reduction in signal comprehension, and thus, one should observe a higher level of regulation on signal production faculties in many biological organisms in which communication is performed by production and comprehension of signals \cite{Salahshour2}. Besides, by identifying two different phases of collective motion (informed and misinformed) separated by a discontinuous transition, we have identified the mechanism by which a fraction of informed individuals able to only make noisy observations of the environment, can lead the group. Finally, we have cast the much-studied order-disorder transition in Vicsek-like models into a broader context by showing that how the nature of this transition depends on the amount of information about the environment, available in the population.
	
	A similar comprehension-production asymmetry in the inference capability and similar phase transitions had been recently observed in a model of collective decision making in a population of immobile agents \cite{Salahshour2,Salahshour3}. In this model, a population of agents who reside on a fixed communication network and live in an environment that can take one out of a finite number of possible states, try to collectively infer the environmental state \cite{Salahshour2,Salahshour3}. Similarly to the environmental state, in this model, individuals can take one out of a finite number of beliefs and are equipped with a production and a comprehension probability transition matrices to produce and comprehend a set of a finite number of signals. It is shown that such a model of collective decision making with signaling shows a similar comprehension-production asymmetry, according to which an amount of noise in signal production is more detrimental for the collective capability of the population to infer the environmental state, compared to the same amount of noise in signal production \cite{Salahshour2}. Our work here extends this study in different ways. In previous models of biological communication by production and comprehension of signals \cite{Salahshour,Salahshour2,Salahshour3}, or more generally, models of a proto-language \cite{Nowak1,Nowak2,Ke,Salahshour4}, individuals try to communicate a finite number of states, which lack a distance measure. In contrast, here we have explored a case when individuals can form beliefs over a continuous space and produce and comprehend a continuous set of signals, which poses a distance measure. In this regard, in the limit of zero movement speed, our model reduces to a model of biological communication in a population of immobile agents, who can form beliefs, and produce and comprehend signals, over a continuous space of possible belief which poses a distance measure. Second, by introducing movement into the model, we have been able to study collective movement in such a communicating population. Our finding here shows that a similar asymmetry in the collective inference capability of biological populations is at work in this case as well. Furthermore, by showing that the production noise also increases the density fluctuations and the probability of group fragmentation, our analysis reveals new ways in which noise in signal production can be more detrimental than noise in the comprehension. 
	
	Finally, we note that the similarity between the phenomenology of the two apparently different systems, that is the model of collective movement introduced here, and the model of collective decision making mentioned before \cite{Salahshour2,Salahshour3,Salahshour4}, can be understood by noting that movement decision involves choosing a direction between a continuous set of possible directions. This observation highlights the similarity of collective movement with collective decision making, which can provide insights into the physics of collective movement such as the nature of the phase transitions observed in such models or the mechanisms by which such populations can optimize their information acquisition capabilities \cite{Salahshour3}, which can be subject to future researches.
	
	\section*{Methods}
	\subsection*{Overview of the model}
	We consider a population of $N$ individuals moving with constant speed $v_0$, in a two dimensional $L\times L$ surface with periodic boundaries. Individuals need to decide about the direction of their motion $b$. In a two dimensional world, this is an angle, lying in the interval $[0,2\pi]$. Individuals decide about their direction of motion based on social information and possibly personal observation. Social information is provided through communication by exchanging signals. In order to communicate, in each time step individuals produce a signal $\sigma$, given their direction of motion $b$ using their production channel $G(\sigma|b)$, and transmit the signal to their neighborhood. The signals are transmitted up to a distance $l$ such that all the individuals located in a circle of radius $l$ centered on the transmitter receive the signal. The receivers comprehend the signal as referring to a direction $r'$ using their comprehension channel $C(r'|\sigma)$. We assume a fraction $h$ of the individuals are informed individuals. In addition to the information received through communication, informed individuals can make noisy observation of the environment. The result of an observation $r$ is determined based on the environmental direction $\epsilon$ using the observation channel $R(r|\epsilon)$. As a result of observation and communication, individuals receive a set of representations $\pmb{r}=\{r,\pmb{r'}\}$. Here, $r$ is a representation reached by personal observation (which exists only for informed individuals), and $\pmb{r'}$ is the set of representations reached through communication. As a representation is supposed to represent the direction of motion in two dimension, these are continuous variables which lie in the interval $[0,2\pi)$. Each individual, $\alpha$, makes a decision about its direction of motion $b_\alpha$, based on its internal state using the decision rule given in eq. (\ref{eq1}). The decision rule in eq. (\ref{eqdecision}) can be thought of as a simple averaging of the representations, in which personal observation is weighted by a self confidence factor $\omega$. The dynamics repeat in the same way for $T$ time steps.
	
	\subsection*{The simulations}
	All the simulations are started with random distribution of positions and directions of motions. The parameter values used in the simulations differ for each figure, and are given in the figure captions. To derive the hysteresis loop in figure (\ref{fig2}.f), we consider a population of size $N=400$, with a fraction of informed individuals equal to $h=\frac{6}{400}$ (other parameter values are presented in the figure caption). Starting with random distribution of positions and directions of motion, the simulation is run for $T=10000$ time steps, after which $h$ is decreased by a value $\frac{1}{400}$, and the procedure is repeated until $h=-\frac{1}{400}$ is reached. As mentioned before, a negative $h$ results from reversing the environmental direction (here $\epsilon=\pi/2$ is changed to $\epsilon=-\pi/2$). Then $h$ is increased again in steps of $\delta h=\frac{1}{400}$, each time after performing the simulation for $T=10000$ time steps, until $h$ ‌reaches its initial value of $h=\frac{6}{400}$. The angular deviation and its error bars for each value of $h$ are calculated based on the last $T=2000$ steps (after the system equilibriates) of the simulation for each $h$ value. 
	
	\section*{Acknowledgment} The authors are indebted to Yasser Roudi for many fruitful discussions and insightful ideas. M. S. was at Sharif University of Technology when this work started. M.S. acknowledges funding from Alexander von Humboldt Foundation in the framework of the Sofja Kovalevskaja Award endowed by the German Federal Ministry of Education and Research during part of this research.
	
	\section*{Author contributions statement}
	
	M.S. designed the research, S.R. contributed new reagents to the research, M.S. performed the research,  M.S.  wrote the paper. All the authors revised the manuscript.
	
	\section*{Additional information}
	The authors declare no conflict of interest.

\def\refVicsek{16}

%\pagebreak
%\onecolumngrid

\setcounter{figure}{0}
\setcounter{equation}{0}
\setcounter{section}{0}

\newcommand{\red}{\textcolor{red}}
\newcommand{\blue}{\textcolor{blue}}

\renewcommand\thesection{SI. \arabic{section}}
\renewcommand\thesubsection{\thesection.\arabic{subsection}}

\renewcommand\thefigure{SI.\arabic{figure}} 
\renewcommand\thetable{SI.\arabic{table}} 
\renewcommand\theequation{SI.\arabic{equation}} 
%\makeatletter
%\def\p@subsection{}
%\makeatother

\clearpage
%\onecolumn
\begin{center}
	{\huge \bf Supplementary Information
		
	%	Collective Movement with Signaling\\
 }
	\vspace{0.4cm}
	
%	Mohammad Salahshour and Shahin Rouhani\\
	%XXXXX	
	%	{}
	%{\small \it XXXX}
	
\end{center}
\vspace{0.4cm}
\flushbottom

\section{Overview of the model}
We consider a population of $N$ individuals moving in a two dimensional $L\times L$ surface with periodic boundaries, with constant speed $v_0$. Individuals need to make decision about the direction of their motion $b$. In a two dimensional world, this is an angle, lying in the interval $[0,2\pi]$. Individuals make decision about their direction of motion based on social information and possibly personal observation. Social information is provided through communication by exchanging signals. In order to communicate, in each time step individuals produce a signal $\sigma$, given their direction of motion $b$ using their production channel $G(\sigma|b)$, and transmit the signal to their neighborhood. The signals are transmitted up to a distance $l$ such that all the individuals located in a circle of radius $l$ centered on the transmitter receive the signal. The receivers comprehend the signal as referring to a direction $r'$ using their comprehension channel $C(r'|\sigma)$. We assume a fraction $h$ of the individuals are informed individuals. In addition to the information received through communication, informed individuals can make noisy observation of the environment. The result of an observation $r$ is determined based on the environmental direction $\epsilon$ using the observation channel $R(r|\epsilon)$. As a result of observation and communication, individuals receive a set of representations $\bm{r}=\{r,\bm{r'}\}$. Here, $r$ is a representation reached by personal observation (which exists only for informed individuals), and $\bm{r'}$ is the set of representations reached through communication. Representing the direction of motion in two dimension, representations are continuous variables which lie in the interval $[0,2\pi)$. Each individual $\alpha$ makes decision based on its internal state using the following decision rule:
\begin{align}
b_{\alpha}=\omega r_{\alpha}+(1-\omega)\frac{\sum_{r'\in \bm{r'}_\alpha}r'}{|\bm{r'}_\alpha|}.
\label{eqdecision}
\end{align}
Here, $b_{\alpha}$ is the belief of the individual $\alpha$ about the favorable direction to which the individual moves, $r_\alpha$ is the observation of individual $\alpha$ in case it has made an observation, $\bm{r'}_\alpha$ is the set of representations individual $\alpha$ has reached by comprehending signals it has received, $|\bm{r'}_\alpha|$ is the number of these representations, and $\omega\in[0,1]$ is the self confidence of individuals. The decision rule in eq. (\ref{eqdecision}) can be thought of as a simple averaging of the representations, in which personal observation is weighted by a self confidence factor $\omega$. The dynamics repeat in the same way for $T$ time steps.

In addition to this basic model, in the following we consider the possibility of noise in the decision making. For this purpose, we consider a case when an individual intends to travel to the direction $b$, due to a noisy implementation of its decision, in practice it travels to the direction $b+\xi$, where $\xi$ is a variable drawn uniformly at random in the interval $[-\eta_A, \eta_A]$. Such a noise is equivalent to the way noise is incorporated in the Vicsek model [\red{\refVicsek}]. In the following, we show our result, concerning comprehension-production asymmetry is valid in the presence of such a noise in decision making as well.

\begin{figure*}[!ht]
	\centering
	\includegraphics[width=1\linewidth, trim = 70 50 90 10, clip,]{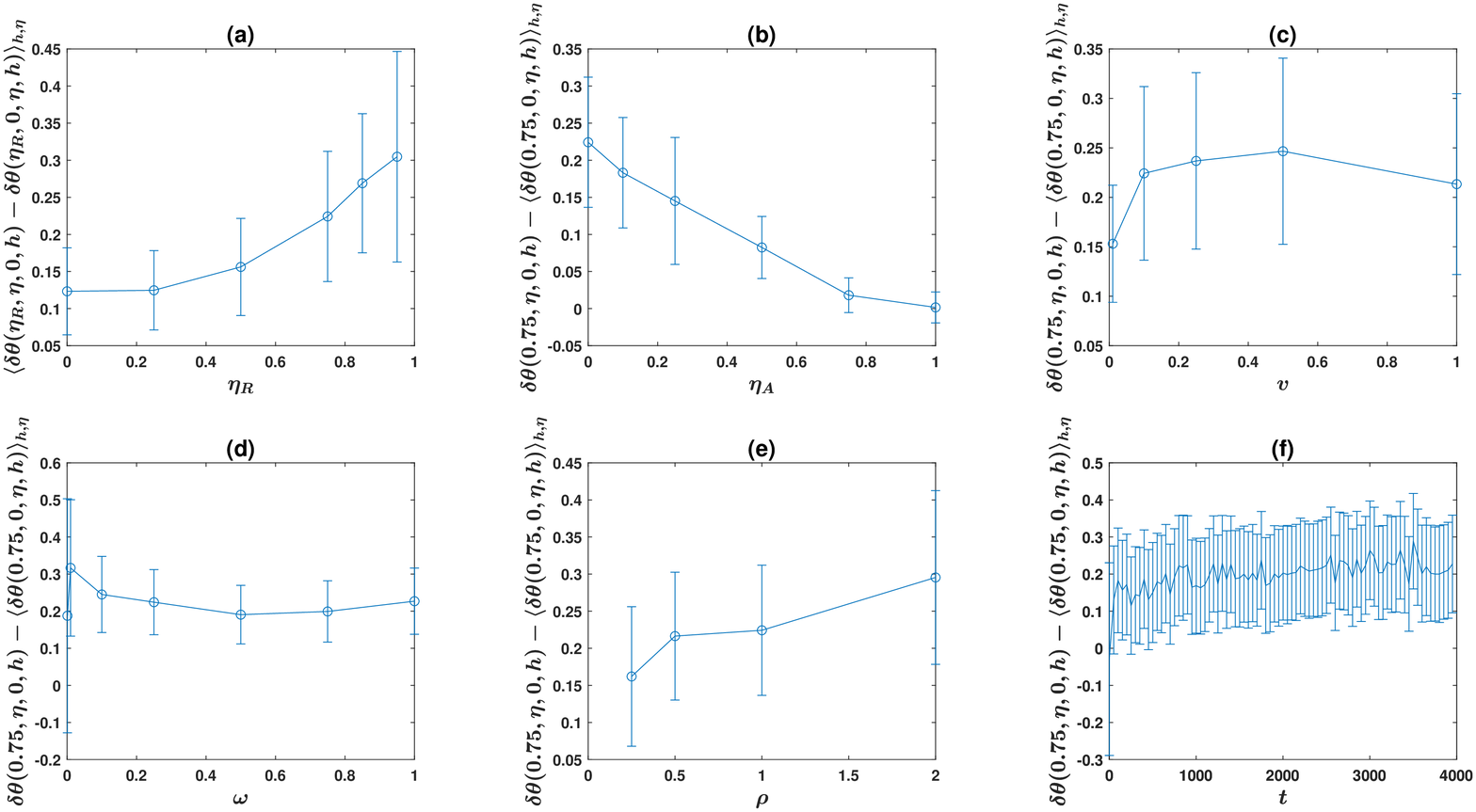}
	\caption{The mean asymmetry in the inference capability over $h$ and $\eta$, $\langle \delta\Theta(\eta_R,\eta,0,h)-\delta\Theta(\eta_R,0,\eta,h)\rangle_{h,\eta}$, as a function of model parameters. The base parameter values used in the simulations are: $N=100$, $v_0=0.1$, $\omega=0.25$, $\rho=1$, $\eta_R=0.75$ and $\eta_A=0$. In each simulation, one of the parameters is changed as specified in the figure, and the mean asymmetry as a function of that parameter is plotted. The simulation is run for $T=4000$ time steps, and a sample of $R=24$ simulations is used. Averages and error bars are calculated based on the last $500$ time steps of the simulation.}
	\label{figasym}
\end{figure*}
\section{The supplementary video}
In the Supplementary Video (SV), a manifestation of the resulting collective motion for the cases that an amount of noise ($\eta=0.25$) is in comprehension (right panel in the SV), and when it is in production (left panel in the SV) is presented. Here, the fraction of informed individuals $h$ is set to $0.1$, the signal decay length $l=1$, the population size is $N=400$, the density $\rho=\frac{1}{16}$ (i.e. the population lives on a two dimensional surface of size $80\times80$), the speed of motion $v_0=0.1$, noise in observation $\eta_R=0.75$, and the self confidence of the individuals is $\omega=0.25$. Here the environmental direction is $\epsilon=\frac{\pi}{2}$ (north).
The video shows with comprehension noise, the population is more successful in collectively finding and traveling towards the environmental direction. In addition, with production noise, the population is strongly fragmented into dense groups, traveling independently of each other. This fact increases the density fluctuations of the population, as seen in the mathematical analysis of the model in the main text, and here, in the Supplementary Information. In addition, the video shows with production noise, informed individuals tend to leave the group, and spend more time traveling alone towards a direction suggested by their own personal information.
\section{Dependence of the asymmetry on model parameters}
\begin{figure*}[!ht]
	\centering
	\includegraphics[width=1\linewidth, trim = 50 20 90 10, clip,]{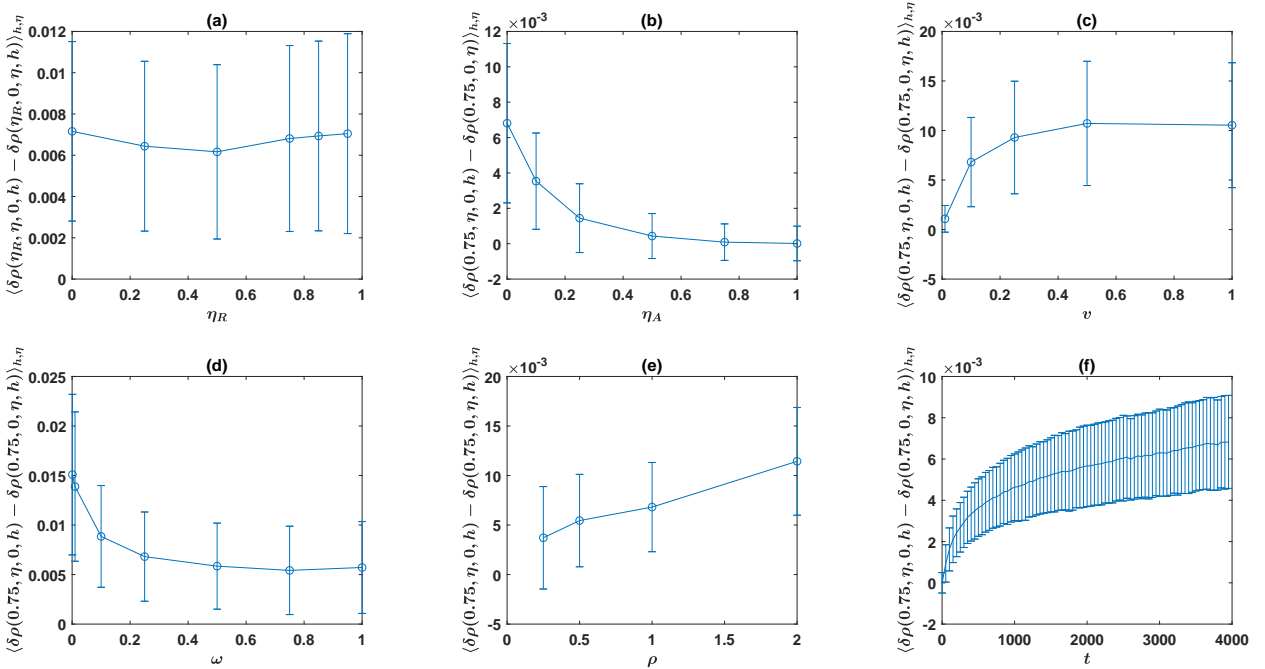}
	\caption{The mean asymmetry in the density fluctuation, over $h$ and $\eta$, $\langle \delta\rho(\eta_R,\eta,0,h)-\delta\rho(\eta_R,0,\eta,h)\rangle_{h,\eta}$, as a function of model parameters. The base parameter values used in the simulations are: $N=100$, $v_0=0.1$, $\omega=0.25$, $\rho=1$, $\eta_R=0.75$ and $\eta_A=0$. In each simulation, one of the parameters is changed as specified in the figure, and the mean asymmetry as a function of that parameter is plotted. The simulation is run for $T=4000$ time steps, and a sample of $R=24$ simulations is used. Reported values and error bars are calculated based on the final state.}
	\label{figasymdense}
\end{figure*}
In this section we show that the asymmetry found in the main text holds for different parameter values of the model, and also investigate how the asymmetry depends on model parameters. For this purpose, we run two simulation sets. We consider an amount of noise $\eta$ and, in one simulation set, put it in the comprehension, taking the production noiseless, and in the second simulation set, put the noise in production, taking the comprehension noiseless. We define the asymmetry in the inference capability as the  difference between the angular deviations of the population in the case that noise is in production, from that when the noise is in comprehension, $\delta\Theta(\eta_R,\eta,0,h)-\delta\Theta(\eta_R,0,\eta,h)$. Where the angular deviation of the population $\delta\theta$, is the (angular) difference between the average direction of the population and the environmental direction. In the same way, the asymmetry in density fluctuations is defined as the difference between density fluctuations when an amount of noise is in production from that when the noise is in comprehension, $\delta\rho(\eta_R,\eta,0,h)-\delta\rho(\eta_R,0,\eta,h)$. We begin by the asymmetry in the inference capability. As shown in the main text this quantity is always positive. In fact, this positivity holds for all the parameter values. To investigate the dependence of the asymmetry on model parameters, we consider the average of asymmetry over $h$ and $\eta$, $\langle \delta\Theta(\eta_R,\eta,0,h)-\delta\Theta(\eta_R,0,\eta,h)\rangle_{h,\eta}$. We run simulations using the base parameter values $N=100$, $v_0=0.1$, $\omega=0.25$, $\rho=1$, $\eta_R=0.75$ and $\eta_A=0$. To see how the asymmetry behaves with respect to model parameters, in each simulation set, we change one of the parameters as specified in the figure, and plot the average asymmetry as a function of the given parameter, in Fig. (\ref{figasym}). The simulations are run for $T=4000$ time steps. The values and their error bars reported in the figure are averages and standard deviations based on the last $T=500$ time steps of the simulations. In addition, an average over a sample of $R=24$ runs is taken.

We begin by investigating the dependence of the asymmetry on $\eta_R$ in Fig. (\ref{figasym}.a). As can be seen, the mean asymmetry is positive for all values of $\eta_R$ and increases with increasing $\eta_R$. The reason is with a noisier sensor, the population relies more on social information acquired through communication to find the environmental direction. As the effect of information provided by communication increases, the asymmetry, being rooted in communication, increases as well.

The effect of a possible noise that individuals may have in decision making is investigated in Fig. (\ref{figasym}.b). As can be seen, the asymmetry decreases with increasing noise in decision making. The reason is that with a noisier decision making, larger amount of the information provided through communication is lost. The asymmetry decreases in the same pace as the information acquisition capability of the population decreases with increasing $\eta_A$.

The effect of speed of travel and self confidence $\omega_R$, are investigated in, respectively Fig. (\ref{figasym}.c), and Fig. (\ref{figasym}.d). Here, it can be seen that the asymmetry is maximized for moderate velocities. It decreases for very small $\omega$, This is because for a very small $\omega$ the population goes to misinformed phase with higher probability, and thus fails to infer the environmental direction for both cases of production and comprehension noise. However, the asymmetry increases for large enough $\omega$ and shows small sensitivity with varying $\omega$ for larger values. 

The effect of density of the population $\rho$, and time $T$, are investigated in, respectively Fig. (\ref{figasym}.e), and Fig. (\ref{figasym}.f). The asymmetry increases with increasing density as can be seen in Fig. (\ref{figasym}.e). Regarding the dependence of asymmetry on time, it can be seen that the asymmetry rapidly increases for very short times, and shows smaller variation with time for larger times.

The dependence of the asymmetry in density fluctuations is investigated in Fig. (\ref{figasymdense}). Here, the same simulations as that presented in Fig. (\ref{figasym}) are used. The simulations are run for $T=4000$ time steps and the asymmetry in density fluctuation, averaged over $h$ and $\eta$, at the final time of simulation is plotted as a function of different model parameters. Values and error bars are calculated based on a sample of $R=24$ simulations. Here, the unnormalized density fluctuation is plotted. To have a criteria of the relative magnitude of the asymmetry in density fluctuations with respect to the absolute value of the density fluctuations, we note that the value of the density fluctuation is of the same order of the asymmetry in the density fluctuation. To save ink and paper, in the following by asymmetry we mean asymmetry in density fluctuation. As can be seen the mean asymmetry over $h$ and $\eta$ is always non-negative. The same is true for asymmetry in each value of $h$ and $\eta$.

In Fig. (\ref{figasymdense}.a) the mean asymmetry in density fluctuation as a function of noise in representation $\eta_R$ is plotted, where it can be seen that the mean asymmetry shows small sensitivity with respect to variations in the representation noise. In Fig. (\ref{figasymdense}.b), the mean asymmetry in density fluctuation as a function of noise in decision making is plotted. As can be seen the asymmetry decreases with $\eta_A$, and tends to zero for large values of $\eta_A$. This is because for large $\eta_A$, the system goes to the disordered phase for all values of $\eta$ for both comprehension and production noise.

In Fig. (\ref{figasymdense}.c), the mean asymmetry in density fluctuation as a function of speed $v_0$ is plotted, where it can be seen that the asymmetry increases with speed. This is because the density fluctuation for production noise increases with speed, while that for comprehension noise shows little variation by increasing the speed.

In Fig. (\ref{figasymdense}.d), we can see that the mean asymmetry decreases with $\omega$. In fact, the asymmetry in density fluctuation is the largest in the misinformed collective motion phase, and decreases in the informed collective motion phase. As increasing $\omega$ decreases the value of $h$ in which transition from misinformed collective motion to informed collective motion occurs, the mean asymmetry in density fluctuation over $h$ and $\eta$ decreases with increasing $\omega$.

In Fig. (\ref{figasymdense}.e), we see that the mean asymmetry in density fluctuations increases with increasing the density. Finally, in Fig. (\ref{figasymdense}.f), the mean asymmetry in density fluctuation as a function of time is plotted. Here we see that the mean asymmetry increases with time. The rate of its increase decreases for larger times, and it eventually equilibrates to a stationary value for large enough times (not shown in the figure). In fact, the density fluctuation for comprehension noise equilibrate to a stationary value rather soon, and the increase in the asymmetry is due to increase in the density fluctuation with production noise with time.


\begin{thebibliography}{99}
		\bibitem{Czirok}Czir\'ok, Andr\'as, Eshel Ben-Jacob, Inon Cohen, and Tam\'as Vicsek. "Formation of complex bacterial colonies via self-generated vortices." Physical Review E 54, no. 2 (1996): 1791.
		\bibitem{Sokolov}Sokolov, Andrey, Igor S. Aranson, John O. Kessler, and Raymond E. Goldstein. "Concentration dependence of the collective dynamics of swimming bacteria." Physical review letters 98, no. 15 (2007): 158102.
		\bibitem{Szabo}Szabo, Balint, G. J. Sz\"oll\"osi, B. G\"onci, Zs Jur\'anyi, David Selmeczi, and Tam\'as Vicsek. "Phase transition in the collective migration of tissue cells: experiment and model." Physical Review E 74, no. 6 (2006): 061908.
		\bibitem{Friedl}Friedl, Peter, and Darren Gilmour. "Collective cell migration in morphogenesis, regeneration and cancer." Nature reviews Molecular cell biology 10, no. 7 (2009): 445.
		\bibitem{Buhl}Buhl, Jerome, David JT Sumpter, Iain D. Couzin, Joe J. Hale, Emma Despland, Edgar R. Miller, and Steve J. Simpson. "From disorder to order in marching locusts." Science 312, no. 5778 (2006): 1402-1406.
		\bibitem{Couzin1}Couzin, Iain D., and Nigel R. Franks. "Self-organized lane formation and optimized traffic flow in army ants." Proceedings of the Royal Society of London. Series B: Biological Sciences 270, no. 1511 (2003): 139-146.
		\bibitem{Ward}Ward, Ashley JW, David JT Sumpter, Iain D. Couzin, Paul JB Hart, and Jens Krause. "Quorum decision-making facilitates information transfer in fish shoals." Proceedings of the National Academy of Sciences 105, no. 19 (2008): 6948-6953.
		\bibitem{Bajec}Bajec, Iztok Lebar, and Frank H. Heppner. "Organized flight in birds." Animal Behaviour 78, no. 4 (2009): 777-789.
		\bibitem{Fischhoff}Fischhoff, Ilya R., Siva R. Sundaresan, Justine Cordingley, Heather M. Larkin, Marie-Jeanne Sellier, and Daniel I. Rubenstein. "Social relationships and reproductive state influence leadership roles in movements of plains zebra, Equus burchellii." Animal Behaviour 73, no. 5 (2007): 825-831.
		\bibitem{Sueur}Sueur, Cedric, and Odile Petit. "Organization of group members at departure is driven by social structure in Macaca." International Journal of Primatology 29, no. 4 (2008): 1085-1098.
		\bibitem{Faria}Faria, Jolyon J., John RG Dyer, Colin R. Tosh, and Jens Krause. "Leadership and social information use in human crowds." Animal Behaviour 79, no. 4 (2010): 895-901.
		\bibitem{Vicsek}Vicsek, Tam\'as, and Anna Zafeiris. "Collective motion." Physics reports 517, no. 3-4 (2012): 71-140.
		\bibitem{Ward2}Ward, Ashley JW, James E. Herbert-Read, David JT Sumpter, and Jens Krause. "Fast and accurate decisions through collective vigilance in fish shoals." Proceedings of the National Academy of Sciences 108, no. 6 (2011): 2312-2315.
		\bibitem{Grunbaum}Grünbaum, Daniel. "Schooling as a strategy for taxis in a noisy environment." Evolutionary Ecology 12, no. 5 (1998): 503-522.
		\bibitem{Berdahl}Berdahl, Andrew M., Albert B. Kao, Andrea Flack, Peter AH Westley, Edward A. Codling, Iain D. Couzin, Anthony I. Dell, and Dora Biro. "Collective animal navigation and migratory culture: from theoretical models to empirical evidence." Philosophical Transactions of the Royal Society B: Biological Sciences 373, no. 1746 (2018): 20170009.
		\bibitem{Vicsek2}Vicsek, Tam\'as, András Czir\'ok, Eshel Ben-Jacob, Inon Cohen, and Ofer Shochet. "Novel type of phase transition in a system of self-driven particles." Physical review letters 75, no. 6 (1995): 1226.
		\bibitem{Chate}Chat\'e, Hugues, Francesco Ginelli, Guillaume Gr\'egoire, Fernando Peruani, and Franck Raynaud. "Modeling collective motion: variations on the Vicsek model." The European Physical Journal B 64, no. 3-4 (2008): 451-456.
		\bibitem{Gregoire}Gr\'egoire, Guillaume, and Hugues Chat\'e. "Onset of collective and cohesive motion." Physical review letters 92, no. 2 (2004): 025702.
		\bibitem{Couzin}Couzin, Iain D., et al. "Effective leadership and decision-making in animal groups on the move." Nature 433.7025 (2005): 513.
		\bibitem{Bricard}Bricard, Antoine, Jean-Baptiste Caussin, Nicolas Desreumaux, Olivier Dauchot, and Denis Bartolo. "Emergence of macroscopic directed motion in populations of motile colloids." Nature 503, no. 7474 (2013): 95.
		\bibitem{Cucker}Cucker, Felipe, and Steve Smale. "Emergent behavior in flocks." IEEE Transactions on automatic control 52, no. 5 (2007): 852-862.
		\bibitem{Nagai}Nagai, Ken H., Yutaka Sumino, Raul Montagne, Igor S. Aranson, and Hugues Chat\'e. "Collective motion of self-propelled particles with memory." Physical review letters 114, no. 16 (2015): 168001.
		\bibitem{Luca}De Luca, Giancarlo, et al. "Fishing out collective memory of migratory schools." Journal of the Royal Society Interface 11.95 (2014): 20140043.
		\bibitem{McCann}McCann, Colin P., Paul W. Kriebel, Carole A. Parent, and Wolfgang Losert. "Cell speed, persistence and information transmission during signal relay and collective migration." J Cell Sci 123, no. 10 (2010): 1724-1731.
		\bibitem{Haas}Haas, Petra, and Darren Gilmour. "Chemokine signaling mediates self-organizing tissue migration in the zebrafish lateral line." Developmental cell 10, no. 5 (2006): 673-680.	
		\bibitem{Rappel}Rappel, Wouter-Jan. "Cell–cell communication during collective migration." Proceedings of the National Academy of Sciences 113, no. 6 (2016): 1471-1473.
		\bibitem{Leonhardt}Leonhardt, Sara Diana, et al. "Ecology and evolution of communication in social insects." Cell 164.6 (2016): 1277-1287.
		\bibitem{Sumpter}Sumpter, David JT. "The principles of collective animal behaviour." Philosophical transactions of the royal society B: Biological Sciences 361, no. 1465 (2005): 5-22.
		\bibitem{Monaco1}Monaco, J.D., Hwang, G.M., Schultz, K.M. and Zhang, K., 2019, May. Cognitive swarming: an approach from the theoretical neuroscience of hippocampal function. In Micro-and Nanotechnology Sensors, Systems, and Applications XI (Vol. 10982, p. 109822D). International Society for Optics and Photonics.
		\bibitem{Monaco2}Monaco, J.D., Hwang, G.M., Schultz, K.M. and Zhang, K., 2020. Cognitive swarming in complex environments with attractor dynamics and oscillatory computing. Biological Cybernetics, 114(2), pp.269-284.
		\bibitem{Salahshour2}Salahshour, Mohammad, Shahin Rouhani, and Yasser Roudi. "Phase transitions and asymmetry between signal comprehension and production in biological communication." Scientific reports 9, no. 1 (2019): 3428.
		\bibitem{Brumm}Brumm, Henrik, ed. Animal communication and noise. Vol. 2. Springer Science \& Business Media, 2013.
		\bibitem{Schuster}Schuster, Sophie, Sue Anne Zollinger, John A. Lesku, and Henrik Brumm. "On the evolution of noise-dependent vocal plasticity in birds." Biology letters 8, no. 6 (2012): 913-916.
		\bibitem{Wiley}Wiley, R. Haven. "How noise determines the evolution of communication." Animal behaviour 124 (2017): 307-313.
		\bibitem{Hotchkin}Hotchkin, Cara, and Susan Parks. "The Lombard effect and other noise‐induced vocal modifications: insight from mammalian communication systems." Biological Reviews 88, no. 4 (2013): 809-824.
		\bibitem{Salahshour}Salahshour, Mohammad, and Shahin Rouhani. "Evolutionary value of collective sensing." arXiv preprint arXiv:1802.03524 (2018).
		\bibitem{Goldenfeld}Goldenfeld, Nigel. "Lectures on phase transitions and the renormalization group." (1992).
		\bibitem{Binder}Binder, Kurt. "Theory of first-order phase transitions." Reports on progress in physics 50.7 (1987): 783.
		\bibitem{Salahshour3}Salahshour, M., 2019. Phase Diagram and Optimal Information Use in a Collective Sensing System. Physical review letters, 123(6), p.068101.
		\bibitem{Nowak1}Nowak, M.A. and Krakauer, D.C., 1999. The evolution of language. Proceedings of the National Academy of Sciences, 96(14), pp.8028-8033.
		\bibitem{Nowak2}Nowak, M.A. and Komarova, N.L., 2001. Towards an evolutionary theory of language. Trends in cognitive sciences, 5(7), pp.288-295.
		\bibitem{Ke}Ke, J., Minett, J.W., Au, C.P. and Wang, W.S.Y., 2002. Self‐organization and selection in the emergence of vocabulary. Complexity, 7(3), pp.41-54.
		\bibitem{Salahshour4}Salahshour, M., 2020. Coevolution of cooperation and language. Physical Review E, 102(4), p.042409.
	\end{thebibliography}
\end{document}